\RequirePackage[hyphens]{url}
\documentclass[sigconf]{acmart}
\usepackage{xspace}
\usepackage{enumitem}
\usepackage{fancyvrb}
\usepackage[group-separator={,}]{siunitx}
\usepackage{svg}
\usepackage{wrapfig}
\usepackage[utf8]{inputenc}
\usepackage{multirow}
\usepackage{pifont}
\usepackage{listings}
\usepackage{xurl}
\AtBeginDocument{%
  }

\copyrightyear{2025}
\acmYear{2025}
\setcopyright{cc}
\setcctype{by}
\acmConference[KDD '25]{Proceedings of the 31st ACM SIGKDD Conference on Knowledge Discovery and Data Mining V.2}{August 3--7, 2025}{Toronto, ON, Canada}
\acmBooktitle{Proceedings of the 31st ACM SIGKDD Conference on Knowledge Discovery and Data Mining V.2 (KDD '25), August 3--7, 2025, Toronto, ON, Canada}
\acmDOI{10.1145/3711896.3737435}
\acmISBN{979-8-4007-1454-2/2025/08}

\title{CURE: A Dataset for Clinical Understanding \& Retrieval Evaluation}

\newcommand\cure{CURE\xspace}
\newcommand\curefullname{Clinical Understanding \& Retrieval Evaluation\xspace}
\newcommand\clinia{Clinia\xspace}

\newcommand{\cmark}{\ding{51}}%
\newcommand{\xmark}{\ding{55}}%

\definecolor{PineGreen}{rgb}{0.0, 0.5, 0.0}

\newcommand\En{\texttt{en}}
\newcommand\Es{\texttt{es}}
\newcommand\Fr{\texttt{fr}}

\newcommand{\insertcollectioncomparison}{
\begin{table*}[ht!]
    \centering
    \Huge
    \resizebox{\textwidth}{!}{%
        \begin{tabular}{@{}lccc|c|cccc@{}}
        \toprule
        \multirow{2}{*}{\textbf{Dataset Name}} & \multirow{2}{*}{\textbf{Query Type}} & \textbf{Human} & \multirow{2}{*}{\textbf{CLIR}} & \multirow{2}{*}{\textbf{Domain}} & \multirow{2}{*}{\textbf{\#Q}} & \textbf{Avg} & \textbf{Total} & \textbf{Corpora} \\
        & & \textbf{Labels} & & & & \textbf{\#Labels/Q} & \textbf{\# Labels} & \textbf{Source} \\
        \midrule
        Nutrition Facts \cite{boteva2016full} & Titles, Descriptions etc. & \xmark$^{[1]}$ & \xmark & Nutrition & 3.8 & 3237 & 12334 & PubMed Abstracts \\
        TREC-COVID \cite{roberts2021searching} & Natural questions & \cmark &  \xmark & COVID & 50 & 493.5 & 24673 & CORD-19 Papers \\        
        SciDocs \cite{cohan2020specter} & Article titles & \xmark$^{[1]}$ & \xmark & Science & 1000 & 29.9 & 29928 \\
        SciFacts \cite{wadden2020fact} & Assertions & \cmark & \xmark & COVID & 1119 & 1.1 & 1258 & CORD-19 Papers \\
        BioASQ \cite{krithara2023bioasq} & Natural questions & \cmark &  \xmark & Biomedicine & 5389 & 9.7 & 52413 & PubMed \\
        \midrule
        \multirow{6}{*}{\textbf{\cure (our work)}} & \multirow{6}{*}{Natural questions$^{[2]}$} & \multirow{6}{*}{\cmark$^{[3]}$} & \multirow{6}{*}{\cmark} & Dentistry \& Oral Health, & \multirow{6}{*}{2000} & \multirow{6}{*}{40.4} & \multirow{6}{*}{80716} \\
        & & & & Dermatology, Gastroenterology, & & & & BioMed Central, \\
        & & & & Genetics, Neuroscience \& & & & & PubMed Central, \\
        & & & & Neurology, Orthopedic Surgery, & & & & and Nature$^{[1]}$ \\
        & & & & Otorhinolaryngology, Plastic Surgery, & & & & \\
        & & & & Psychiatry \& Psychology, and Pulmonology & & & & \\
        \bottomrule
        \end{tabular}%
    }
    \caption{Comparison of related retrieval collections. \textbf{Human Labels} indicates human-generated labels (vs. synthetically generated labels); \textbf{CLIR} indicates whether the collection supports cross-lingual retrieval; \textbf{Avg \# Labels/Q} is the average number of labels provided per query; \textbf{Total \# Labels} is the total number of human labels (both positive and negative) across all splits. $^1$Nutrition Facts \& SciDocs construct queries from article titles; $^2$We translate human-generated queries; $^3$We also annotate query-passage pairs using LLMs as described in \autoref{sec:reljudg}.}
    \label{tab:collection_comparison}
\end{table*}
}

\newcommand{\insertsampleannotation}{
    \begin{table}[t]
    \centering
    \Huge
    \resizebox{\columnwidth}{!}{%
        \begin{tabular}{p{0.15\textwidth}p{0.85\textwidth}}
        \hline
        \toprule
        Query: & \textbf{What is the treatment for iatrogenic facial nerve neuroapraxia after facelifting procedure?} \\
        Passage: & As for the cases of facial nerve neuroapraxia, they were treated by massage of affected muscles, physiotherapies in the form of galvanic stimulation, corticosteroids, and vitamin B12 injection for 8 weeks until complete recovery, which was usually achieved between 3 weeks and 3 months. \\
        \bottomrule
        \hline
        \end{tabular}
    }
    \caption{Example of a generated \textit{query} and identified \textit{relevant passage} from the domain of Plastic Surgery}
    \label{tab:sampleannotation}
    \end{table}
}

\newcommand{\insertsamplequeries}{
    \begin{table}[t]
    \centering
    \Huge
    \resizebox{\columnwidth}{!}{%
        \begin{tabular}{p{0.15\textwidth}p{0.85\textwidth}}
        \hline
        \toprule
        Layman: & \textit{What is the most common \textcolor{PineGreen}{location of a fracture in facial trauma} associated with dental injuries?} \\
        Expert: & \textit{Is there any \textcolor{PineGreen}{functionality enabled by Erich arch bars} that IMF screws cannot provide?} \\
        \bottomrule
        \hline
        \end{tabular}
    }
    \caption{Example of \textit{layman} query and \textit{expert} query in the domain of Dental and Oral Health}
    \label{tab:samplequeries}
    \end{table}
}

\newcommand{\insertsamplefeedback}{
    \begin{table}[t]
    \centering
    \Huge
    \resizebox{\columnwidth}{!}{%
        \begin{tabular}{p{0.20\textwidth}p{0.85\textwidth}}
        \hline
        \toprule
        Query: & \textbf{What is the impact of physical activity on reducing the risk of gastric cancer, compared to other prevention methods?} \\
        \vspace{2pt}
        Passage: & In this condition, seeking prevention methods is key to reducing the burden of gastric cancer. This covers tactics including expanding access to healthcare and screening programs, encouraging healthy lifestyles, and minimizing exposure to established risk factors like smoking and H.pylori infection [35]. One of the easy-accessible methods in this regard is physical activity. \\
        \vspace{2pt}
        Feedback: & \textcolor{red}{\textit{The passage is not relevant with respect to the query, or at least, there is no statement on the different impact of physical activity compared to other methods}} \\
        \bottomrule
        \hline
        \end{tabular}
    }
    \caption{Example of annotation with 
    \textit{feedback} provided during validation}
    \label{tab:samplefeedback}
    \end{table}
}

\newcommand{\insertsamplerelevancejudgment}{
    \begin{table}[t]
    \centering
    \Huge
    \resizebox{\columnwidth}{!}{%
        \begin{tabular}{p{0.20\textwidth}p{0.85\textwidth}}
        \hline
        \toprule
        Query: & \textbf{Is treatment of class II malocclusion recommended during childhood?} \\
        \vspace{2pt}
        Passage: & The best time and approach to treat Class II malocclusion in one or two stages has been extensively investigated in Orthodontics. 
        Systematic reviews in the literature provide the best scientific evidence, and one of the sources for them is the Cochrane Library . One of these reviews published at Cochrane Library concluded that early treatment of Class II malocclusion and maxillary incisor protrusion in adolescence was not more efficient than treatment in a single stage in the beginning of puberty . Randomized clinical trials conducted in the United States and England , also concluded
        that, after the second stage of the treatment of Class II malocclusion, there were no differences between the group submitted to early intervention
        and the group treated in a single stage later.
        Therefore, the choice of a two- or one-stage treatment seems to be a matter of professional preference, rather than a biological decision. \\
        Judgment: & \textit{2} \\
        Explanation: & \textcolor{red}{\textit{The query asks whether treatment of Class II malocclusion is recommended during childhood. The passage provides a detailed review of the literature, including systematic reviews and randomized clinical trials, which conclude that early treatment of Class II malocclusion in childhood (adolescence) is not more efficient than treatment in a single stage during puberty. The passage clearly indicates that the choice between early and late treatment is a matter of professional preference rather than a biological necessity. This information directly addresses the query by suggesting that early treatment is not necessarily recommended over later treatment, providing a clear and actionable answer.}} \\
        \bottomrule
        \hline
        \end{tabular}
    }
    \caption{Example of a \textit{relevance judgment} with \textit{explanation}}
    \label{tab:samplerelevancejudgment}
    \end{table}
}

\newcommand{\insertdescriptivestats}{
    \begin{table*}[t]
    \centering
    \small
    \resizebox{\textwidth}{!}{%
        \begin{tabular}{lcccccccccc}
        \toprule
            \multirow{2}{*}{\textbf{Domain}} & \multirow{2}{*}{\textbf{\# Passages}} & \multirow{2}{*}{\textbf{\# Articles}} & \multirow{2}{*}{\textbf{\# Queries}} & \multirow{2}{*}{\textbf{\# Judgements}} & \multicolumn{3}{c}{\textbf{Avg. Q Len}} & \multirow{2}{*}{\textbf{Avg. P Len}} \\
            &&&&& \texttt{en} & \texttt{es} & \texttt{fr} \\
            \midrule
            Dentistry \& Oral Health & 29352 & 5436 & 200 & 7966 & 11.6 & 13.6 & 14.6 & 75.0 \\
            Dermatology & 15687 & 3751 & 200 & 2692 & 8.6 & 10.1 & 11.2 & 74.3 \\
            Gastroenterology & 23068  & 5832 & 200 & 10665 & 12.0 & 13.8 & 14.9 & 73.9 \\
            Genetics & 26996  & 6581 & 200 & 13605 & 11.2 & 13.9 & 12.9 & 87.6 \\
            Neuroscience \& Neurology & 30388 & 6245 & 200 & 8409 & 12.0 & 13.9 & 16.1 & 79.2 \\
            Orthopedic Surgery & 24910 & 4663 & 200 & 3002 & 13.3 & 15.7 & 16.6 & 76.3 \\
            Otorhinolaryngology & 24192 & 4424 & 200 & 7342 & 12.4 & 14.8 & 15.6 & 70.7 \\
            Plastic Surgery & 26605 & 4995 & 200 & 5085 & 12.6 & 14.9 & 16.3 & 70.6 \\
            Psychiatry \& Psychology & 35756 & 5967 & 200 & 13753 & 13.5 & 15.9 & 16.6 & 85.1 \\
            Pulmonology & 32317 & 5562 & 200 & 8197 & 12.6 & 15.2 & 15.8 & 74.8 \\
            \midrule
            & 244600 & 51083 & 2000 & 80716 & 12.0 & 14.1 & 15.2 & 77.2 \\
        \bottomrule
        \end{tabular}
    }
    \caption{Descriptive statistics for all domains in \cure. \textbf{\# Avg. Q Len:} average number of tokens per query for English (\texttt{en}), Spanish (\texttt{es}), \& French (\texttt{fr}); \textbf{\# Avg. P Len:} average number of tokens per passage. Tokens are delimited by whitespace.}
    \label{tab:descriptivestats}
    \end{table*}
}

\newcommand{\insertadditionalbaselineresults}{
    \normalsize
    \begin{table}[h!]
        \centering
        \normalsize
        \resizebox{\columnwidth}{!}{%
        \begin{tabular}{l|ccc|ccc}
            \toprule
            & \multicolumn{3}{c|}{NDCG@10} & \multicolumn{3}{c}{Recall@100} \\
            Model Name & \En-\En & \Fr-\En & \Es-\En & \En-\En & \Fr-\En & \Es-\En \\
            \midrule 
            paraphrase-multilingual-MiniLM-L12-v2 \cite{reimers-2019-sentence-bert} & 0.309 & 0.203 & 0.225 & 0.481 & 0.337 & 0.366 \\
            
            multilingual-e5-small \cite{wang2024multilingual} & 0.504 & 0.202 & 0.163 & 0.648 & 0.332 & 0.281 \\
            
            multilingual-e5-base \cite{wang2024multilingual} & 0.553 & 0.257 & 0.194 & 0.697 & 0.407 & 0.340 \\
            
            snowflake-arctic-embed-m-v2.0 \cite{yu2024arcticembed20multilingualretrieval} & 0.559 & 0.485 & 0.491 & 0.715 & 0.632 & 0.646 \\
            
            snowflake-arctic-embed-l-v2.0 \cite{yu2024arcticembed20multilingualretrieval} & 0.573 & 0.518 & 0.522 & 0.714 & 0.650 & 0.657 \\

            jina-embeddings-v3 \cite{sturua2024jinaembeddingsv3multilingualembeddingstask} & 0.484 & 0.428 & 0.441 & 0.656 & 0.581 & 0.607 \\
            
            google-text-multilingual-embedding-002 \cite{lee2024geckoversatiletextembeddings} & 0.510 & 0.395 & 0.403 & 0.689 & 0.576 & 0.591 \\

            cohere-embed-multilingual-v3.0 \cite{coheremultilingualembeddings} & 0.587 & 0.522 & 0.527 & 0.730 & 0.576 & 0.663 \\
                        
            openai-text-embedding-3-small \cite{openai2024embeddings} & 0.520 & 0.430 & 0.427 & 0.697 & 0.611 & 0.614 \\
            
            openai-text-embedding-3-large \cite{openai2024embeddings} & \textbf{0.598} & \textbf{0.559} & \textbf{0.554} & \textbf{0.751} & \textbf{0.712} & \textbf{0.713} \\
            \bottomrule
        \end{tabular}%
        }
    \caption{Additional results for various multilingual dense embedders on \cure.}
    \label{tab:AdditionalBaselineResults}
    \end{table}
}

\newcommand{\inserttranslationresults}{
    \normalsize
    \begin{table}[h!]
        \centering
        \normalsize
        \resizebox{\columnwidth}{!}{%
        \begin{tabular}{l|c|c|c|c|c|c|}
            \toprule
            Model Name & Accuracy & Clarity & Fluency & Appropriateness & Proper Names & Grammar \\
            \midrule 
            Mistral-7b-Instruct \cite{jiang2023mistral7b} & 1.1  & 1.133 & 1.2 & 1.833 & 0.967 & 1.3\\
            Bing Microsoft \cite{microsoftMicrosoftTranslator} & 1.767 & 1.7 & 1.567 & 1.933 & 1.733 & 1.733 \\
            ChatGPT4-turbo-2024-04-09 \cite{OpenAI_ChatGPT4_Turbo_2024} & 1.933 & 1.933 & 1.8 & 1.9 & 1.767 & 1.967 \\
            Reverso \cite{reversoReversoCorporate} & 1.833 & 1.733 & 1.767 & 1.833 & 1.6 & 1.867 \\
            Facebook-m4t-v2-large \cite{seamless2023} & 1.6 & 1.567 & 1.667 & 1.933 & 1.467 & 1.833 \\
            Llama-3-8b-instruct \cite{llama3modelcard} & 1.6 & 1.433 & 1.367 & 1.933 & 1.233 & 1.467 \\
            DeepL \cite{deeplDeepLTranslate} & 1.933 & 1.933 & 1.833 & 1.967 & 1.867 & 1.967 \\
            \bottomrule
        \end{tabular}%
        }
    \caption{Evaluation Results for Translation Tools}
    \label{tab:TranslationResults}
    \end{table}
}
\definecolor{lightgreenpastel}{RGB}{204, 255, 204}

\newcommand{\insertpiechart}{
\begin{figure}[h]
{\includegraphics[width=\columnwidth]{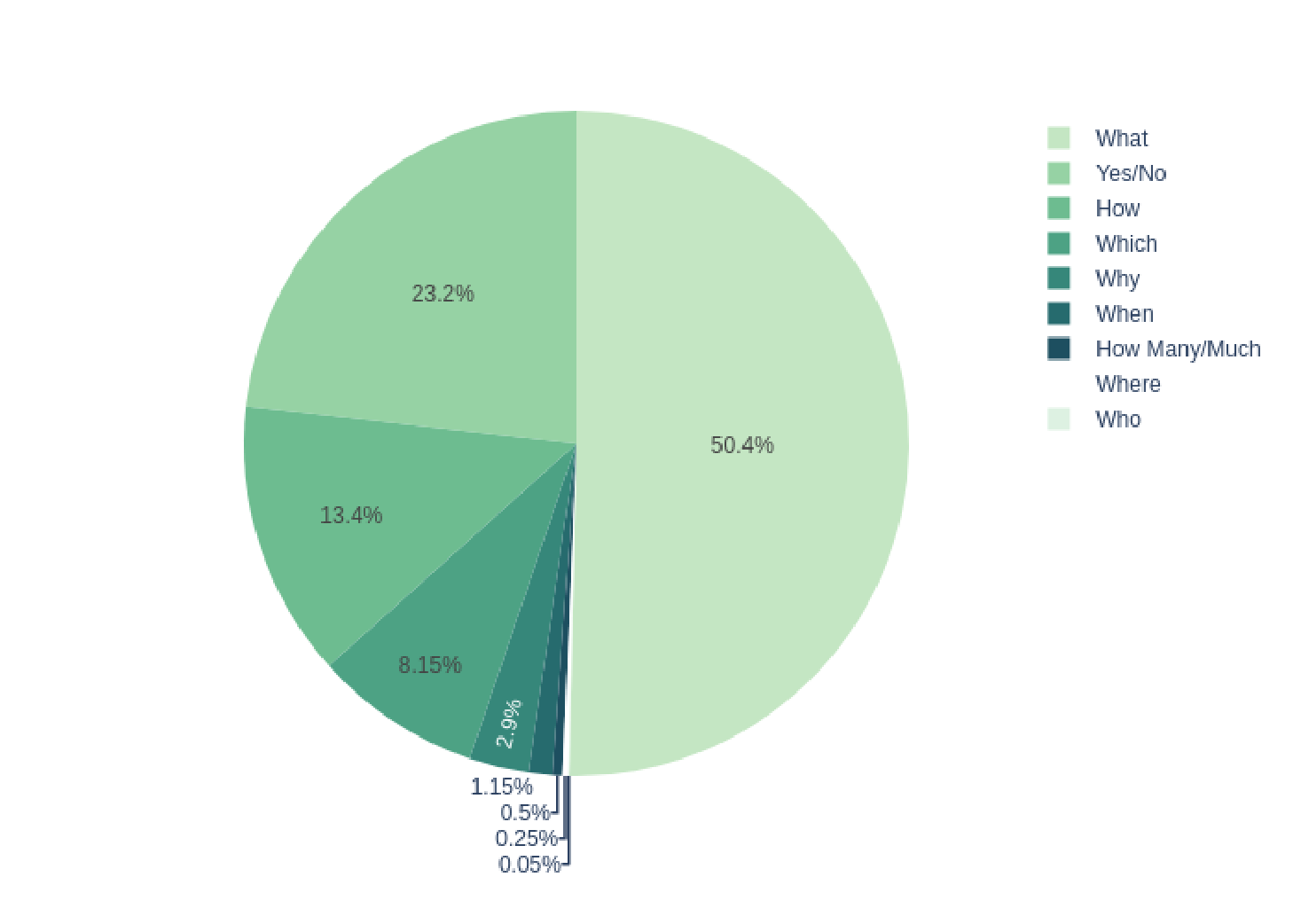}}
\caption{Query Type Distribution across the domains in \cure}\
\label{fig:piechart}
\end{figure}
}

\newcommand{\insertargillascreenshot}{
    \begin{figure}[h!]
    \centering
    \includegraphics[width=\columnwidth]{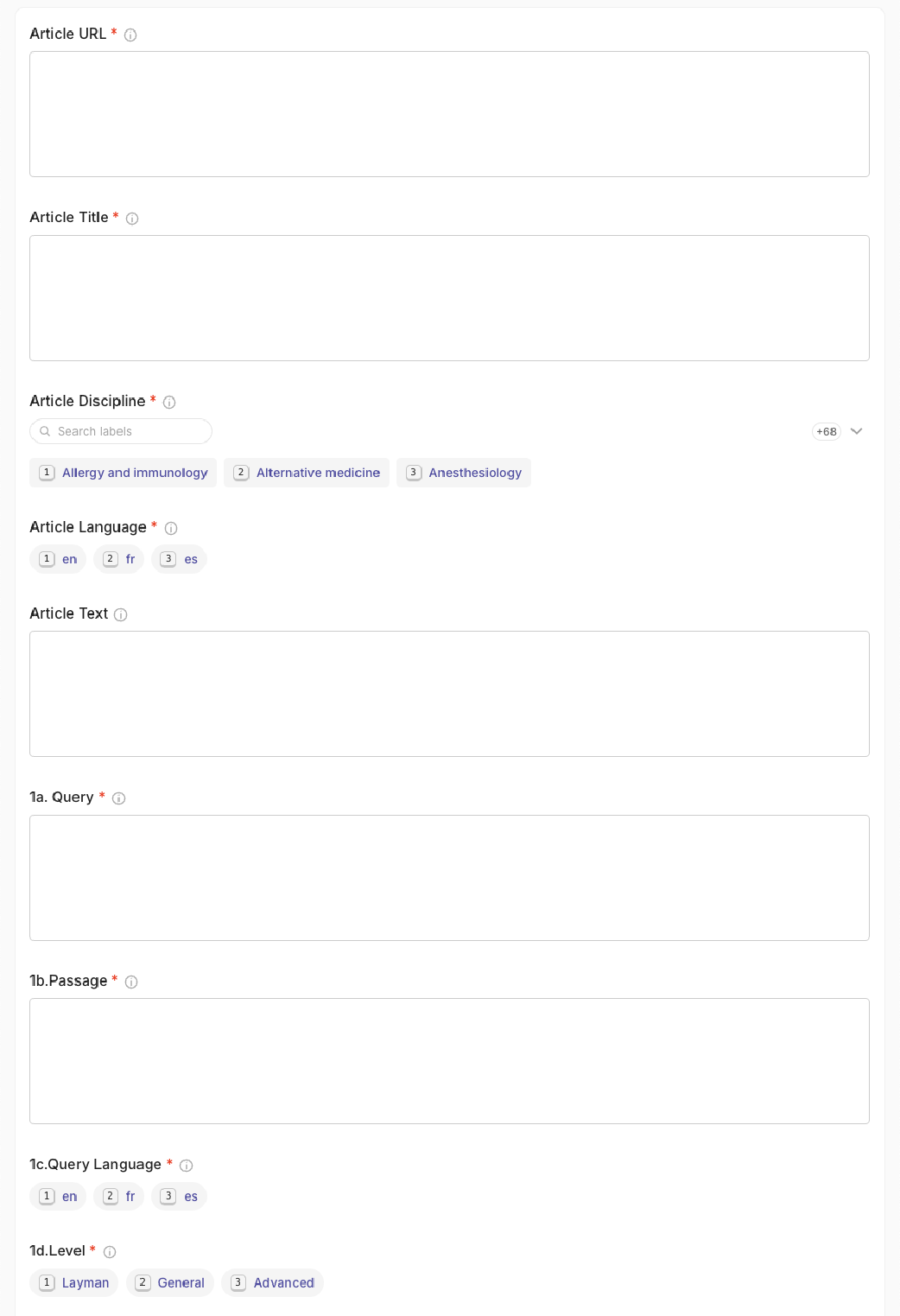}
    \caption{Screenshot of Argilla Interface used for Annotation Generation}
    \label{fig:RelevantPassagesPerQuery}
    \end{figure}
}

\makeatletter
\renewenvironment{quote}
  {\list{}{\rightmargin=2em \leftmargin=2em \topsep=1em}
   \item\relax\itshape}
  {\endlist}
\makeatother

\begin{document}



\author{Nadia Athar Sheikh}
\authornote{Both authors contributed equally to this research.}
\email{nadia.sheikh@clinia.com}
\author{Daniel Buades Marcos}
\authornotemark[1]
\email{daniel.buades@clinia.com}
\affiliation{%
  \institution{Clinia}
  \city{Montreal}
  \state{Quebec}
  \country{Canada}
}

\author{Anne-Laure Jousse}
\affiliation{%
  \institution{Clinia}
  \city{Montreal}
  \state{Quebec}
  \country{Canada}
}
\email{anne-laure.jousse@clinia.com}

\author{Akintunde Oladipo}
\affiliation{%
  \institution{Clinia}
  \city{Montreal}
  \state{Quebec}
  \country{Canada}
}
\email{akintunde.oladipo@clinia.com}

\author{Olivier Rousseau}
\affiliation{%
  \institution{Clinia}
  \city{Montreal}
  \state{Quebec}
  \country{Canada}
}
\email{olivier.rousseau@clinia.com}

\author{Jimmy Lin}
\affiliation{%
  \institution{University of Waterloo}
  \city{Waterloo}
  \state{Ontario}
  \country{Canada}
}
\email{jimmylin@uwaterloo.ca}

\renewcommand{\shortauthors}{Nadia Athar Sheikh et al.}

\begin{abstract}
Given the dominance of dense retrievers which do not generalize well beyond their training dataset distributions, domain-specific test sets are essential in evaluating retrieval performance. Few test datasets are available for retrieval systems intended for use by healthcare providers in a point-of-care setting. To fill this gap we have collaborated with medical professionals to create \cure, a test dataset for passage retrieval composed of 2,000 expert written queries spanning 10 medical domains with a monolingual (English) and two cross-lingual (French/Spanish to English) conditions. In this paper, we describe how \cure was constructed and provide baseline results to showcase its effectiveness as an evaluation tool. \cure is published with a CC BY-NC 4.0 license and can be accessed on \href{https://huggingface.co/datasets/clinia/CUREv1}{Hugging Face}\footnote{https://huggingface.co/datasets/clinia/CUREv1} and as a retrieval task on \href{https://github.com/embeddings-benchmark/mteb}{MTEB}.
\end{abstract}

\begin{CCSXML}
<ccs2012>
   <concept>
       <concept_id>10002951.10003317</concept_id>
       <concept_desc>Information systems~Information retrieval</concept_desc>
       <concept_significance>500</concept_significance>
       </concept>
   <concept>
       <concept_id>10002951.10003317.10003371.10010852</concept_id>
       <concept_desc>Information systems~Environment-specific retrieval</concept_desc>
       <concept_significance>500</concept_significance>
       </concept>
   <concept>
       <concept_id>10002951.10003317.10003359.10003360</concept_id>
       <concept_desc>Information systems~Test collections</concept_desc>
       <concept_significance>500</concept_significance>
       </concept>
   <concept>
       <concept_id>10002951.10003317.10003359.10003362</concept_id>
       <concept_desc>Information systems~Retrieval effectiveness</concept_desc>
       <concept_significance>500</concept_significance>
       </concept>
   <concept>
       <concept_id>10002951.10003317.10003359.10011699</concept_id>
       <concept_desc>Information systems~Presentation of retrieval results</concept_desc>
       <concept_significance>100</concept_significance>
       </concept>
 </ccs2012>
\end{CCSXML}

\ccsdesc[500]{Information systems~Information retrieval}
\ccsdesc[500]{Information systems~Environment-specific retrieval}
\ccsdesc[500]{Information systems~Test collections}
\ccsdesc[500]{Information systems~Retrieval effectiveness}
\ccsdesc[100]{Information systems~Presentation of retrieval results}

\keywords{Biomedical information retrieval, Cross-lingual retrieval, Test datasets, Point-of-care, Dense retrievers, Sparse retrievers}

\maketitle

\section{Introduction}

Evidence-based medicine requires point-of-care healthcare professionals to stay up to date with advances in their specialties to make data-driven decisions when providing care to patients \cite{Sackett1997}. Despite this imperative, research findings have been estimated to take up to 17 years to fully implement in practice \cite{Balas2000-af}.

An impediment for practitioners is the increasingly rapid rate at which relevant literature is published \cite{KamtchumTatuene2021}. Each year, more than one million new articles are published in biomedicine and life sciences \cite{GonzlezMrquez2024}. 

Given the ever-expanding landscape of biomedical information, applications that support ad hoc retrieval are essential tools for health care professionals to provide evidence-based care \cite{Lu2011}. Using ad hoc retrieval, a healthcare professional can express their information need as a query, initiating a search that returns likely relevant documents from a vast corpus \cite{Lu2011}.

Although the vast majority of published biomedical literature is in English, healthcare providers may practice in other languages \cite{Baethge2008}. To effectively accommodate this linguistic diversity, tools aimed at efficient biomedical information access should enable cross-lingual retrieval. 

Information retrieval systems and techniques are typically evaluated using test datasets. There are a wide variety of general test datasets available for evaluating retrieval. At its inception, MTEB (Massive Text Embedding Benchmark) made available 15 retrieval test sets, a collection that continues to expand \cite{muennighoff2022mteb}. To our knowledge, there are no cross-lingual datasets suitable for evaluating retrieval systems intended for use by healthcare providers in a point-of-care setting. As such, domain-specific test sets are essential given the dominance of dense retrievers, whose performance does not generalize well to out-of-domain tasks \cite{thakur2021beir, Ren2023}.

At \clinia, we build products that power search and discovery in health workflows, and the absence of such test datasets has hampered product development, as we cannot benchmark developed solutions systematically. To address this gap, we have introduced a domain-specific test set, which we have released under an open-source license.

In this work, we introduce \curefullname\xspace (\cure ), a test dataset to evaluate ad hoc retrieval systems for use by medical practitioners in a point-of-care setting. We present results from competitive baselines to demonstrate the utility of \cure as a test dataset and to enable comparison with future systems.

\cure is a retrieval dataset with a monolingual and two cross-lingual conditions, with splits spanning ten medical domains. Queries in \cure are natural language questions formulated by health care providers. They express the information needs of practitioners consulting academic literature in the course of their duties. Queries are available in English, French and Spanish. The corpus is constructed by mining an index of English passages extracted from biomedical academic articles. We have made this dataset available to spur research and development on retrieval in a point-of-care setting. 

\section{Dataset Construction}
 \cure is a test collection for passage retrieval. In passage retrieval, given a corpus \textit{C} of \textit{n} passages and a query \textit{q}, the objective is to retrieve a list of \textit{k} passages, sorted by decreasing relevance. A relevant passage meets the information need expressed by the query. In the context of a dataset, the relevance of a passage may be marked on a binary (e.g. Relevant or Not Relevant) or graded (e.g. Perfectly Relevant, Partially Relevant, Related, Irrelevant) scale. In \cure, passages are marked as Relevant, Partially Relevant, and Not Relevant for each query. 

\cure includes 2,000 queries evenly distributed over 10 splits corresponding to the following domains: Dentistry and Oral Health, Dermatology, Gastroenterology, Genetics, Neuroscience \& Neurology, Orthopedic Surgery, Otorhinolaryngology, Plastic Surgery, Psychiatry \& Psychology, and Pulmonology. Each split contains $50$ layman and $150$ expert level queries.

In addition to identified relevant passages, \cure's corpus includes passages mined from an index of more than $7$M biomedical passages. These were extracted from articles whose publishers allowed their use for evaluation. Sources include the \href{https://pmc.ncbi.nlm.nih.gov/tools/openftlist/}{PMC Open Access Subset} \cite{pmc2003a} as well as articles scraped from \href{https://www.nature.com/}{Nature} and \href{https://www.biomedcentral.com/}{BioMedCentral}. Articles were parsed using the \href{https://corpus.tools/wiki/Justext}{JusText} parser and segmented into passages using the parsed document structure. Low-quality passages were filtered out by employing a selection of heuristics introduced to clean training data for Gopher \cite{https://doi.org/10.48550/arxiv.2112.11446}. The corpus contains 244,600 passages from 51,083 articles. The distribution of passages and articles across domains is presented in \autoref{tab:descriptivestats}.

\cure was built in three phases: Annotation, Query Translation, and Corpus Construction. 

\begin{itemize}
    \item During the annotation phase, medical practitioners expressed their information needs as queries and identified relevant passages in the literature they selected. The annotations were then validated and processed.
    \item During query translation, queries were translated from the language in which the annotators had originally written them into English (\En), French (\Fr) and Spanish (\Es) as needed.
    \item The corpus was constructed using relevant passages identified by human annotators and by mining an index of passages extracted from biomedical articles. Large candidate pools were identified for each query and a Large Language Model (LLM) was used to annotate passages for relevance.
\end{itemize} 

 After corpus construction, English queries were used to construct the monolingual conditions of \cure, while Spanish and French queries were used to construct cross-lingual conditions.

\subsection{Annotation}

To ensure high-quality annotations, \cure was constructed with medical professionals. The annotation team included physicians and nurses who were recruited from our professional networks, as well as native speakers of French, English, and Spanish. Annotators were compensated for their work. To ensure consistency and quality, the annotations were validated and processed. Annotation was carried out over 9 months by 12 annotators.

\paragraph{Generation}
To create a test dataset with queries representative of the information needs of point-of-care medical practitioners, we asked annotators to formulate queries based on their own information needs and identify relevant passages from biomedical articles they selected to address those needs. Annotators were instructed to express their queries as natural language questions in English, French, or Spanish, choosing the language they felt most comfortable with. Annotators identified journals and articles they deemed appropriate for their specialties and needs. Before annotation, we ensured that publishers permitted the use of these articles for evaluation.

Annotators were provided with a detailed set of guidelines with accompanying examples to help them with their task. Below is a condensed version for reference:

\begin{quote}
Using a selected scientific article of your choice, generate a query-passage pair and evaluate its audience.
\begin{itemize}
    \item The passage should either be or contain the answer to the query and should be an excerpt from the article. Do not paraphrase.
    \item The passage should be a single paragraph or a short list.
    \item If a single query is answered by multiple passages, please generate distinct query-pairs for each passage.
    \item Identify the article by providing its URL and its title.
    \item If available, provide the full text of the article.
\end{itemize}
\end{quote}

An example of a generated question with the corresponding identified relevant passage is presented in \autoref{tab:sampleannotation}.

\insertsampleannotation

\href{https://argilla.io/}{Argilla} \cite{argilla} was used to collect annotations. The annotation interface is presented in \autoref{appendixA}. Annotators provided the query text and identified relevant passages. Additionally, they specified the query's language, discipline(s), and level of expertise. Annotators identified queries as layman or expert. Layman queries represent those asked by medical practitioners seeking to refresh their knowledge or communicate with non-experts, while expert queries reflect those posed by specialized medical practitioners. An example of a layman and expert query is presented in \autoref{tab:samplequeries}.

\insertsamplequeries

\paragraph{Validation}

Each annotation was reviewed by lead annotators to ensure high quality. In evaluating generated queries, they considered the following criteria:

\begin{itemize}[noitemsep]
\item Is the question likely to be asked by a health professional? 
    \begin{itemize}[noitemsep]
        \item Does it sound natural?
        \item Is it medically relevant?
        \item Is it an impactful clinical question?
    \end{itemize}
\item Does the question match the level of expertise with which it is associated?
\item Is the question well written, considering syntax, fluency, and the presence of typos?
\item Is the question specific enough or too broad?
\end{itemize}

The identified relevant passages were also validated to ensure that they were self-contained and met the information need expressed by the query.

If annotations did not meet the required criteria, the validators provided feedback and corrected annotations were re-evaluated before inclusion in the dataset. Across all annotations, 12.4\% of annotations were flagged for revision. The most common sources of error were related to formulation. An example of an invalid annotation with the corresponding feedback is presented in \autoref{tab:samplefeedback}. 
\insertsamplefeedback

\paragraph{Processing}

Annotators may have entered slight variations of the same passage as relevant for distinct queries as a passage may meet more than one information need. Additionally, since relevant passages were entered manually, variations in text may have occurred. Furthermore, a passage that was identified as relevant for one query may have been embedded in a passage identified as relevant for another query. Although not identified as such, the containing passage would still be relevant for the query associated with the contained passage.

In processing annotations, minor variants of the same passage were identified automatically using Levenshtein distance between passages as a heuristic and the longest passage among variants was selected to replace variants. In addition, passages containing these variations were identified and marked as partially relevant for the query associated with the contained passage. 

\subsection{Query Translation}

Queries were translated so that each query is available in English, French, and Spanish. To select the best-performing translation tool, we sampled a set of queries and ran them through various candidate tools. Then, three native-speaking medical annotators were asked to rate each translation from 0 (not meeting criteria) to 2 (fully meeting criteria) based on how well they met each of the following criteria: 

\begin{itemize}
    \item \textit{Accuracy}: The content should faithfully convey the message and sentiment of the original text in the target language. No added or forgotten content.
    \item \textit{Clarity}: The message must be easily understandable, and any instructions should be actionable and easy to follow.
    \item \textit{Fluency}:	Does the translation flow naturally in the target language?		
    \item \textit{Appropriateness}: Consider cultural nuances and ensure that the translation is appropriate for the target audience. Check for register, idiomatic expressions and cultural references that may need adaptation.
    \item \textit{Proper names and terms}: Are medical terms, names, trademarks, and other non-translatable words preserved from the source text?								
    \item \textit{Grammar and style}: Check for grammatical correctness in the translated text. Evaluate whether the translator has captured the style and tone of the original content.													
\end{itemize}

Averaged results across annotators and queries for each tool are presented in \autoref{tab:TranslationResults}
\inserttranslationresults

In selecting a tool, the criteria \textit{accuracy} and \textit{proper names and terms} were given the highest weights, followed by \textit{clarity} and \textit{fluency}. The criteria of \textit{appropriateness} and \textit{grammar and style} were given the lowest weights. \href{https://hnd.www.deepl.com/en/products/translator}{DeepL API} \cite{deeplDeepLTranslate} was selected based on its strong performance relative to alternative tools, and we used it to translate the remaining queries. Finally, native speakers performed quality checks and made manual modifications to ensure that the resulting translations were of high quality.

\subsection{Corpus Construction}

Our corpus was constructed by mining for in-domain passages and assigning relevance judgments to passages from large candidate pools using an LLM.

\paragraph{Mining}

Following \citet{10.1162/tacl_a_00595}, we mined passages using Reciprocal Rank Fusion (RRF) combining KNN and BM25 retrievers. Iterating through all validated queries, a search was executed using each query and one relevant passage with each type of retriever. Embeddings for the KNN retrievers were generated using \texttt{BGE-small-v1.5} \cite{bge_embedding}. Each KNN retriever considered 250 passages from each shard and returned 50 passages. The BM25 retriever using the query to search employed a match query with a slop of 2 while the BM25 retriever using the relevant passage to search employed a match phrase query permitting fuzzy matches.
Source articles contributing to the top 50 passages returned were considered to be in-domain and their passages were included in \cure's corpus.

\paragraph{Candidate Identification}
In any large retrieval dataset, manually annotating all query-passage pairs for relevance is not feasible. Typically, small sets of candidates are selected for annotation while the remaining passages are presumed to be irrelevant. As Thakur et al. noted, in constructing BEIR, since only small candidate pools are annotated for relevance, datasets can be biased. For example, Thakur et al. identified a bias in TREC-COVID likely arising from the use of lexical cues to construct candidate pools \cite{thakur2021beir}.

To mitigate this bias, we generated large candidate pools for each query using the GTE multilingual cross-encoder \cite{zhang2024mgtegeneralizedlongcontexttext}. The generated queries and associated passages were assigned to distinct splits, each corresponding to one of ten medical domains. Queries and passages belonging to multiple domains were placed in a single split. The English variant of each query was scored against all passages within the split, and the pool for each query was constructed by taking passages with a score above 0.42. This cutoff score was established using a test set of passages annotated by humans for relevance and was found to maximize the difference between the rate of relevant passages and non-relevant passages included in the candidate pool.

\paragraph{Relevance Judgments}
\label{sec:reljudg}

The passages in the candidate pools were annotated for relevance using Qwen 2.5 $72$B \cite{qwen2}. The use of LLMs for passage annotation was inspired by Upadhyay et al.'s work, which found a strong correlation between human relevance judgments and LLM-derived judgments \cite{https://doi.org/10.48550/arxiv.2405.04727}. Qwen 2.5 was selected because of its relatively strong concordance with human annotations on a test set. We used a variation of the prompt from Upadhyay et al presented in \autoref{appendixB}. Generated explanations are included in the dataset. An example explanation is presented in \autoref{tab:samplerelevancejudgment}\insertsamplerelevancejudgment. To evaluate the quality of relevance judgments in our final dataset, we sampled a set of query-passage pairs from our dataset and had two annotators independently assess relevance. Their judgments were compared with those in our dataset, yielding a Fleiss’ Kappa of 0.691.

After constructing the corpus, query translations were used to generate a monolingual (English) and two cross-lingual conditions (Spanish-English, French-English) of the dataset with conditions differing only in query language. 

\section{Dataset Statistics}
 
The average number of tokens per query and passages in the corpus is presented in table \autoref{tab:descriptivestats}.

The distribution of question types across the dataset is presented in \autoref{fig:piechart}
As can be seen, 50.35\% of all queries are \textit{What} questions, followed by 13.45\% of queries which are \textit{Yes/No} questions. In addition to factoid questions, 13.45\% of \cure queries are \textit{How} questions, and 2.9\% are \textit{Why} questions.
On average, each query is associated with 9.93 perfectly relevant passages and 30.43 partially relevant passages. 

\insertpiechart
\insertdescriptivestats

\section{Retrieval Baselines}

To demonstrate the effectiveness of our test dataset for retrieval, we present results from four different baselines. BM25 serves as our representative sparse retrieval baseline. For a dense retrieval baseline, we choose MedEmbed Base \cite{balachandran2024medembed} as a representative of a domain-specific model, GTE Multilingual Base \cite{zhang2024mgtegeneralizedlongcontexttext} as our representative for models with fewer than 500M parameters, and NV Embed V2 as our representative for models with more than 1B parameters \cite{lee2024nv}.

The results are presented in \autoref{tab:BaselineResults}. We used the MTEB evaluation framework \cite{muennighoff2022mteb} to compute the results. As expected, the largest dense model, NV-Embed, outperforms the smaller dense models, GTE Multilingual Base and MedEmbed Base, which in turn outperforms the sparse BM25 model. Across all retrievers, the performance under monolingual conditions is better than the model under cross-lingual conditions of \cure. 

%\insertbaselineresults
% \newcommand{\insertbaselineresults}{
\begin{table}[h!]
    % \centering
    \resizebox{\columnwidth}{!}{%
    \begin{tabular}{ l|ccc|cccccc}
        \toprule
        & \multicolumn{3}{c|}{NDCG@10}& \multicolumn{3}{c}{Recall@100}\\
        Model Name & \En-\En & \Fr-\En & \Es-\En & \En-\En & \Fr-\En & \Es-\En\\
         \midrule 
        BM-25 \cite{bm25s} & 0.355 & - & - & 0.523 & - & - \\
        MedEmbed-base-v0.1 \cite{balachandran2024medembed} & 0.515 & 0.193 & 0.144 & 0.690 & 0.368 & 0.308 \\
        mGTE Base (304M) \cite{zhang2024mgtegeneralizedlongcontexttext} & 0.558 & 0.528 & 0.481 & 0.714 & 0.623 & 0.643 \\
       NV Embed v2 (7B) \cite{lee2024nv} & \textbf{0.651} & \textbf{0.603} & \textbf{0.604} & \textbf{0.786} & \textbf{0.738} &  \textbf{0.740} \\
       % \bottomrule
       \midrule
    \end{tabular}%
}
% \vspace{-2.0em}
\caption{Sparse and dense baseline results on \cure. }
\label{tab:BaselineResults}
\end{table}
% }

There are many compelling retrieval models available online, making it challenging to select just a few for presentation. As an additional reference, results for a subset of well-known multilingual dense embedders, both open source and proprietary, are provided in \autoref{tab:AdditionalBaselineResults}.

For readers interested in evaluating a broader range of models, a comprehensive and up-to-date overview of additional results is available on the \href{https://huggingface.co/spaces/mteb/leaderboard}{MTEB Leaderboard} \cite{muennighoff2022mteb}. Results specific to \cure can be viewed by selecting the \textit{MTEB(Medical, v1)} benchmark and filtering by task.

The \textit{MTEB(Medical, v1)} benchmark, which not only includes \cure but also features tasks from medical settings distinct from point-of-care, was proposed by Clinia to facilitate the evaluation of new models. As such, Clinia remains committed to maintaining it by continuously evaluating and uploading results for new models as they become available.

\insertadditionalbaselineresults

\section{Related Test Datasets}

In this section, we review widely used biomedical test datasets considering their utility in evaluating ad hoc retrieval systems intended for healthcare professionals. 

Unlike \cure, many commonly used test datasets for evaluating retrieval do not reflect the wide range of concerns of medical professionals. Datasets such as TREC-COVID \cite{roberts2021searching} and SciFacts \cite{wadden2020fact} have a narrow focus, with corpora drawn from CORD-19, a collection of scientific articles on COVID-19 \cite{roberts2021searching}. The Nutrition Facts Corpus dataset \cite{boteva2016full} is similar, as its corpus is constructed from academic texts in the field of nutrition. Other datasets are too broad, extending beyond the field of medicine. Scifacts \cite{wadden2020fact}, for example, has a corpus that includes articles drawn from a wide range of scientific disciplines. BioASQ \cite{krithara2023bioasq}, as used in BEIR, \cite{thakur2021beir} has a domain focus more similar to \cure, with a corpus constructed from PubMed/MedLine articles. However, queries  in BioASQ are broadly representative of the information needs of biomedical scientists, rather than exclusively those of healthcare providers \cite{krithara2023bioasq}. In contrast, the queries in \cure were generated by medical professionals and explicitly validated as being typical of those asked by medical providers in a point of care setting. Of the queries generated, 97.63\% of the queries generated for \cure were identified by 2 or more out of 3 annotators as  being aligned with the types of questions likely to be asked by a health practitioner seeking research articles in a point-of-care setting.

\insertcollectioncomparison

Furthermore, many biomedical retrieval datasets are repurposed from datasets originally intended to evaluate systems on downstream tasks other than retrieval. As a consequence, queries may not be in forms representative of those posed by actual users. For example, in the Nutrition Facts Corpus dataset \cite{boteva2016full}, queries are constructed from article titles and descriptions. The SciDocs dataset \cite{wadden2020fact}, originally designed to test citation prediction and repurposed for retrieval evaluation in BEIR \cite{wadden2020fact}, uses titles of papers as queries. Similarly, the Scifacts dataset, as used in BEIR \cite{wadden2020fact}, originally designed to evaluate fact-checking, uses assertions rather than interrogatives as queries. In contrast, \cure includes queries in natural language formulated by healthcare professionals. 

Relevance judgments in some test retrieval datasets are also limited. For example, in both SciDocs \cite{wadden2020fact} and Nutrition Facts Corpus \cite{boteva2016full} relevance is inferred from citations rather than explicitly judged. In \cure, we include explicit human relevance judgments and mitigate the bias introduced by relatively shallow candidate pools by using an LLM to fill in missing relevance judgments.

\section{Limitations and Future Work}

\cure currently covers a limited number of medical domains. As performance in one domain may not be indicative of performance in another, in future iterations we would like to expand \cure's coverage. We would also like to further diversify the type of queries represented in \cure, with a greater focus on queries that require reasoning, particularly within the context of clinical decision-making. 

Although \cure supports the evaluation of search in academic literature, in a point-of-care setting a medical practitioner may consult a range of different types of documents, including educational materials or Electronic Health Records. In the future, we would like to develop test datasets that enable the evaluation of retrieval systems across different types of documents. 

\section{Conclusion}

In this paper, we present \cure, a test dataset to evaluate retrieval systems in healthcare settings. \cure was built with a team of healthcare professionals, including nurses and doctors, who generated questions and identified relevant passages from biomedical literature. Its corpus was augmented by mining an index of $7$M passages extracted from biomedical articles. To mitigate bias, large candidate pools were constructed and an LLM was used to annotate passages for relevance. The final test dataset consists of 2,000 queries evenly distributed over 10 medical domains and is available in a monolingual (English) and two cross-lingual conditions (French/Spanish to English). We also present the performance of different retrievers on this dataset as baselines. 

There are few, if any, test datasets designed to evaluate retrieval systems intended for use by medical practitioners in a point-of-care setting. At \clinia, we use \cure to benchmark our existing retrieval solutions and guide their development. In making \cure freely available for evaluation, we aim to foster research and the development of reliable retrieval tools for healthcare practitioners.

\begin{acks}

We want to highlight our extensive team of annotators, particularly Dr. Arturo Vela Lasagabaster, MSc, MD; Dr. Leila Cattelan, MD; Dr. Natalia Cullell, MD, PhD; and Mr. Noah Tremblay Taillon for their meticulous work. We also extend our gratitude to our CTO, Etienne Soulard-Geoffrion, and the rest of Clinia’s executive team for their belief in responsible ML development and their support for this initiative.

\end{acks}

\bibliographystyle{ACM-Reference-Format}
\bibliography{custom}

\newpage
\appendix

\section{Annotation Interface}
\label{appendixA}
A screenshot of the interface used to collect annotations is presented in \autoref{fig:RelevantPassagesPerQuery}

\insertargillascreenshot

% \appendix
\vfill\null
\section{Prompt used for Relevance Judgments}\label{appendixB}
\begin{lstlisting}
You are a { clinical_role }, who consults academic journals to ensure that you provide care that is in accordance with best practices. Here, you will be presented a query you constructed along with a retrieved passage from a scientific article. Your task is to assign a score on an integer scale of 0 to 2 based on the degree to which the passage contains an actionable answer to your query:

2: Assign a score of 2 if the passage is dedicated to the query and contains the exact answer
1: Assign a score of 1 if the passage has some answer for the query but the answer is a bit unclear, or hidden amongst extraneous information
0: Assign a score of 0 if the passage seems related to the query but does not answer it OR if the passage has nothing to do with the query

Note: The article title is provided to help you contextualize the passage.

# Important Instruction

Split this task into steps:
1. Identify the underlying information need of the search from the perspective of a trained { profession } seeking to provide informed patient care.
2. Measure how well the passage satisfies the information need expressed by the query. Carefully consider details in both the query and the passage.
3. Before scoring the passage, write a careful explanation justifying your score that includes a detailed rationale.
4. Provide each score in the format of a JSON with the following structure:

{{
    "explanation": "Careful explanation justifying your score that includes a detailed rationale",
    "score": "Your final assigned score based on the explanation. One of: 0, 1, 2"
}}

-------

Query: { query }
Passage: { passage }
Article Title: { article_title }

# Examples

{query_examples}
 
\end{lstlisting}
\end{document}